\documentclass[aps,prl,showpacs,floatfix,twocolumn]{revtex4}
\usepackage{graphicx}
\usepackage{dcolumn}
\usepackage{bm}
\usepackage{stmaryrd}
\usepackage{latexsym}
\usepackage{amssymb}
\usepackage{amsfonts}
\usepackage{amsmath}

\newcommand{\Hm}{\mathcal{H}}
\newcommand{\nf}{n_\mathrm{F}}
\newcommand{\ua}{\uparrow}
\newcommand{\da}{\downarrow}

\begin{document}
\title{Spectral signatures of the Fulde-Ferrell-Larkin-Ovchinnikov  order parameter in one-dimensional optical lattices}
\author{M.~Reza Bakhtiari, M.J.~Leskinen}
\affiliation{Department of Engineering Physics, P.O.Box 5100, 02015 Helsinki University of Technology, Finland}
\author{P.~T\"orm\"a}
\email{paivi.torma@hut.fi}
\affiliation{Department of Engineering Physics, P.O.Box 5100, 02015 Helsinki University of Technology, Finland}

\begin{abstract}
We address an imbalanced two-component atomic Fermi gas restricted by a one-dimensional (1D) optical lattice and an external harmonic potential, 
within the mean-field Bogoliubov-de Gennes formalism. We show that characteristic features of the Fulde-Ferrell-Larkin-Ovchinnikov state are visible in the RF-spectra and in the momentum resolved photoemission spectra of the gas. Specially, Andreev states or mid-gap states can be clearly resolved, which gives a direct experimentally observable signature of the oscillating order parameter. 
\end{abstract}
\maketitle

Experimental realization of spin-density imbalanced Fermi gases~\cite{Zwierlein2006a,Partridge2006a,Giorgini2007a} 
has opened exciting new possibilities to study pairing in systems where matching 
of spin-resolved Fermi surfaces, which is the base of Bardeen-Cooper-Shreffer (BCS) theory, is not valid. Either 
phase separation or extension of the BCS pairing to some other, exotic mechanism is inevitable. The question is of interest in context of various solid state materials~\cite{Radovan2003a,Bianchi2003a,Kakuyanagi2005a,Correa2007a} as well as in hight-energy and astrophysics~\cite{Alford2007a}. One of the main candidates for the non-BCS pairing is the so-called  Fulde-Ferrell-Larkin-Ovchinnikov (FFLO) state~\cite{Fulde1964a,Larkin1964a}. The zero temperature ground state properties of the FFLO state in the context of 1D Fermi gases have been studied extensively within mean-field, exact, and DMRG approaches~\cite{FFLO_papers}. The rapid development of experiments on Fermi gases in optical lattices~\cite{Kohl2005a,Chin2006a,Rom2006a,Jordens2008a}  
suggests that such systems will be available soon. However, it is not obvious how to observe e.g. the spatially varying 
order parameter and other characteristics of the FFLO state, although noise correlations have been proposed to provide information about the pairing correlations~\cite{FFLO_noise}. In this letter we show how the characteristics of the FFLO state are prominently reflected in RF-spectroscopy~\cite{RF_experiments} and in the recently introduced~\cite{Stewart2008a} photoemission spectroscopy of Fermi gases. Unlike the BCS, the FFLO state allows population of single-particle excitations even at zero temperature, corresponding to unpaired particles. For a spatially non-uniform order parameter, some of these excitations can be understood as Andreev bound states residing close to the nodes of the order parameter. We show that such excitations produce distinct features in the spectra, at negative energies, and thus provide a signature of oscillations of the order parameter that is easily distinguishable from the usual pairing signatures at positive energies. Furthermore, by
calculating spectra also at finite temperatures, we show that such features are uniquely related to oscillations of the order parameter. 
 
We consider a two-component attractive Fermi gas confined by an external potential in a 1D lattice with $L$ sites. At low filling, this corresponds to a one-dimensional gas without the lattice. Our qualitative results concerning the spectral signatures should be valid in this case as well. 
We apply a mean-field Bogoliubov-de Gennes (BdG) approach. In 1D, long range order is absent and the mean-field approximation is not as well valid as in 3D, especially the value of the mean-field order parameter may deviate from the exact value~\cite{Marsiglio1997a}. However, mean-field approaches are often applied since they provide qualitative information on the system, as well as make it possible to study the effect of finite temperature and to take into account non-trivial confining geometries (which is not possible in the case of exact analytical solutions). Here, the mean-field description further allows to calculate the RF-spectrum as well as to understand and interpret it in a transparent way. 

At the mean-field level, the system is described by the single-band Hubbard-Hamiltonian:
$
\Hm=-t\,\sum_{i,\sigma}\Big( \hat{c}^\dagger_{i \sigma}\, \hat{c}_{i+1 \sigma}+ \mathrm{h.c}\Big)
+\sum_i \Big(\Delta_i \hat{c}^\dagger_{i \ua}\hat{c}^\dagger_{i \da}+\mathrm{h.c}\Big)  
+\sum_{i\sigma} \Big((V^\mathrm{ext}_i-\mu_\sigma)-U\bar{n}_{i\bar{\sigma}}\Big)\,\hat{n}_{i\sigma} .
$
Here the spin labels $\sigma=\ua,\da$ refer to two hyperfine states of the atoms, the pairing gap is defined as $\Delta_i\equiv-U\langle  \hat{c}_{i \da}\,\hat{c}_{i \ua}\rangle,$ where $U (>0)$ denotes the on-site attractive interaction, $t$ is the hopping strength 
and $V^\mathrm{ext}_i=V_0(i-L/2)^2$ is an external harmonic potential. The $\bar{\sigma}$ 
refers to opposite spin component of $\sigma$. We neglect the Hartree interaction term $U\bar{n}_{i\bar{\sigma}}$ because doing so does not significantly affect the results but simplifies the numerical calculation. We use the Bogoliubov transformation, $\hat{c}_{i \sigma}=\sum_\alpha(u_{\alpha i \sigma}\,\hat{\gamma}_{\alpha\sigma}-\sigma v^*_{\alpha i \sigma}\,\hat{\gamma}_{\alpha\bar{\sigma}}^\dagger)$
to diagonalize the Hamiltonian $\Hm$, leading to the BdG equation
\begin{equation*}
\sum_{j=1}^L \left(
\begin{array}{ccc}
\mathcal{H}^\sigma_{i\,j} & \Delta_{i\,j}  \\
\\
\Delta_{i\,j} & -\mathcal{H}^{\bar\sigma}_{i\,j} \\
\end{array} \right)\,
\left(
\begin{array}{c}
u_{\alpha j \sigma} \\
\\
v_{\alpha j \bar\sigma} \\
\end{array} \right)=E_{\alpha\sigma}\,\left(
\begin{array}{c}
u_{\alpha i \sigma} \\
\\
v_{\alpha i \bar\sigma} \\
\end{array} \right) .
\end{equation*}
Here $\mathcal{H}^\sigma_{i\,j}=-t\,\delta_{i,j\pm1}+\left(V_i^{\mathrm{ext}}-\mu_\sigma\right)\delta_{i\,j}$ is the single particle Hamiltonian and
$\Delta_{i\,j}=\Delta_i \,\delta_{i\,j} $ denotes the local pairing gap. The above equation is solved together with the self-consistency conditions 
for the gap and the spin-resolved densities: 
$
 n_i^\sigma=\sum_{\alpha=1}^L\Big[\,|u_{\alpha i \sigma}|^2\,\nf(E_{\alpha\,\sigma})+|v_{\alpha i \sigma}|^2\,\nf(-E_{\alpha\,\bar\sigma})\,\Big]$,
$
\Delta_i=-U\,\sum_{\alpha=1}^L\Big[\,u_{\alpha i \uparrow}\,v_{\alpha i \downarrow}\nf(E_{\alpha\,\uparrow})-
u_{\alpha i \downarrow}\,v_{\alpha i \uparrow}\nf(-E_{\alpha\,\downarrow})\Big] .
$
Here the densities are normalized to $N_\ua$ and $N_\da$, and $\nf$ is the Fermi function at the temperature $T$.

In RF spectroscopy, the internal state of one of the components ($\uparrow$ or $\downarrow$, energy $\omega_{\scriptscriptstyle{\ua/\da}}$) 
is coupled to a third internal state (denoted final state, energy $\omega_f$) by an RF pulse. The number of particles transferred from the initial to the (initially empty) final state can be observed and gives the spectrum as a function of the detuning, $J_{\scriptscriptstyle{\ua/\da}}(\delta)$, where $\delta = \omega_{RF}-(\omega_f-\omega_{\scriptscriptstyle{\ua/\da}})$. Recently, first experiments where the final state 
momentum was resolved were performed~\cite{Stewart2008a}, in analogy to photoemission spectroscopy of solid state systems. The corresponding spectrum is then $J_{\scriptscriptstyle{\ua/\da}}(\delta,k)$ where $k$ is 
the final state momentum. We have derived $J_{\scriptscriptstyle{\ua/\da}}(\delta,k)$ and $J_{\scriptscriptstyle{\ua/\da}}(\delta)$ in the present case, following the approach~\cite{Torma2000a,Kinnunen2004a}. This corresponds to the quasiparticle picture~\cite{Leskinen2008a} where the RF field can be understood to create quasiparticle excitations by pair breaking rather 
than to induce coherent rotation of the internal states. It is valid when the final state interactions are negligible~\cite{Yu2006a,Schunck2008a}. The spectra become: 
$
J_{\scriptscriptstyle{\ua/\da}}(\delta)= \sum_{K=1}^L J_{\scriptscriptstyle{\ua/\da}}(\delta,K)$ and
\begin{eqnarray}\label{spectra} 
J_{\scriptscriptstyle{\ua/\da}}(\delta,K)=&-&2\pi\sum_{\alpha=1}^L\Big[\,\Big|\sum_{i=1}^L v_{\alpha i \scriptscriptstyle {\da/\ua} }\, 
v_{\scriptscriptstyle{K} i \scriptscriptstyle{\ua\da}}^{\mathrm{\scriptscriptstyle non}}\Big|^2\,\nf(-E_{\alpha\scriptscriptstyle{\da/\ua}}) \nonumber \\
&\times &\delta(E_{\alpha\scriptscriptstyle{\da/\ua}}+\epsilon_K-\delta-\mu_{\scriptscriptstyle{\ua/\da}} )   \nonumber \\
&+&\Big|\sum_{i=1}^L u_{\alpha i \scriptscriptstyle {\ua/\da} }\, 
v_{\scriptscriptstyle{K} i \scriptscriptstyle{\ua\da}}^{\mathrm{\scriptscriptstyle non}}\Big|^2\,\nf(E_{\alpha\scriptscriptstyle{\ua/\da}}) \nonumber \\
& \times&\delta(E_{\alpha\scriptscriptstyle{\ua/\da}}-\epsilon_K+\delta+\mu_{\scriptscriptstyle{\ua/\da}})\Big] .
\end{eqnarray}
Here the symbol $\ua/\da$ denotes the spin-resolved case, while $\ua\da$ and ``non'' refer to the balanced, non-interacting case. The particles in the final state are 
assumed to be trapped in the same potential, therefore the final state wavefunctions in the overlap integrals, 
$v_{\scriptscriptstyle{K} i \scriptscriptstyle{\ua\da}}^{\mathrm{\scriptscriptstyle non}}$, are calculated by solving the BdG equations in the non-interacting case, for high filling in order to have all relevant wavefunctions $v_{\scriptscriptstyle{K} \scriptscriptstyle{\ua\da}}^{\mathrm{\scriptscriptstyle non}}$ non-zero. 
The final state dispersion is also obtained from such calculation:
$\epsilon_K=\mu_{\scriptscriptstyle{\ua\da}}-E_{\scriptscriptstyle{K \da\ua}}$. 
In the numerical calculations, we replace the $\delta$-function by a Lorentzian distribution $\mathcal{L}(x)=\frac{1}{\pi}\frac{\Gamma}{x^2+\Gamma^2}$ and
use the line width $\Gamma=0.05$. All energies are in units of $t$, and we use a realistic trapping potential $V_0=5\times 10^{-4}$ and the attractive interaction strength $U=-3$.

Note that in Eq.(\ref{spectra}) $K$ is the quantum number of the eigenstate of the combined lattice plus harmonic potential. In the limit of vanishing harmonic potential, 
it approaches the lattice momentum $k$. In an experiment, it is the lattice momentum that can be resolved from absorption images after the time of flight expansion. 
Therefore we calculate the observable momentum-resolved spectrum 
by taking into account the overlap between the eigenfunctions for the quantum number $K$ and those corresponding to the lattice momentum $k$:  
$
J_{\scriptscriptstyle{\ua/\da}}(\delta,k)=\sum_{K=1}^L \Big|\sum_{i=1}^L e^{\imath(k i)} 
v_{\scriptscriptstyle{K} i \scriptscriptstyle{\ua\da}}^{\mathrm{\scriptscriptstyle non}}\Big|^2 J_{\scriptscriptstyle{\ua/\da}}(\delta,K) .
$
Effectively, this describes the averaging by the external harmonic potential of the momentum resolved spectra. 

\begin{figure}
\begin{center}
\tabcolsep=0cm
\begin{tabular}{c}
\includegraphics[width=0.77\linewidth]{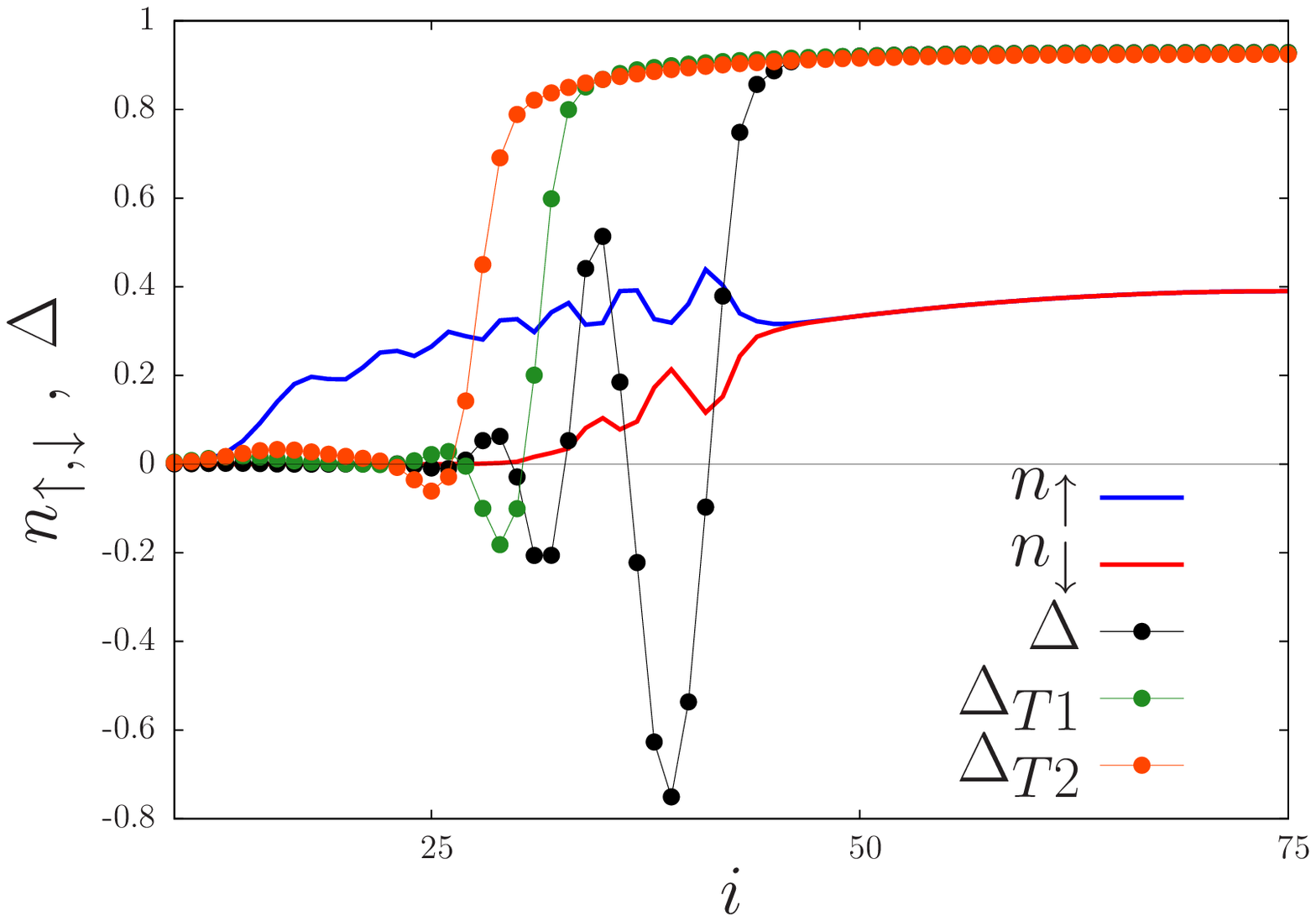}\\
\includegraphics[width=0.8\linewidth]{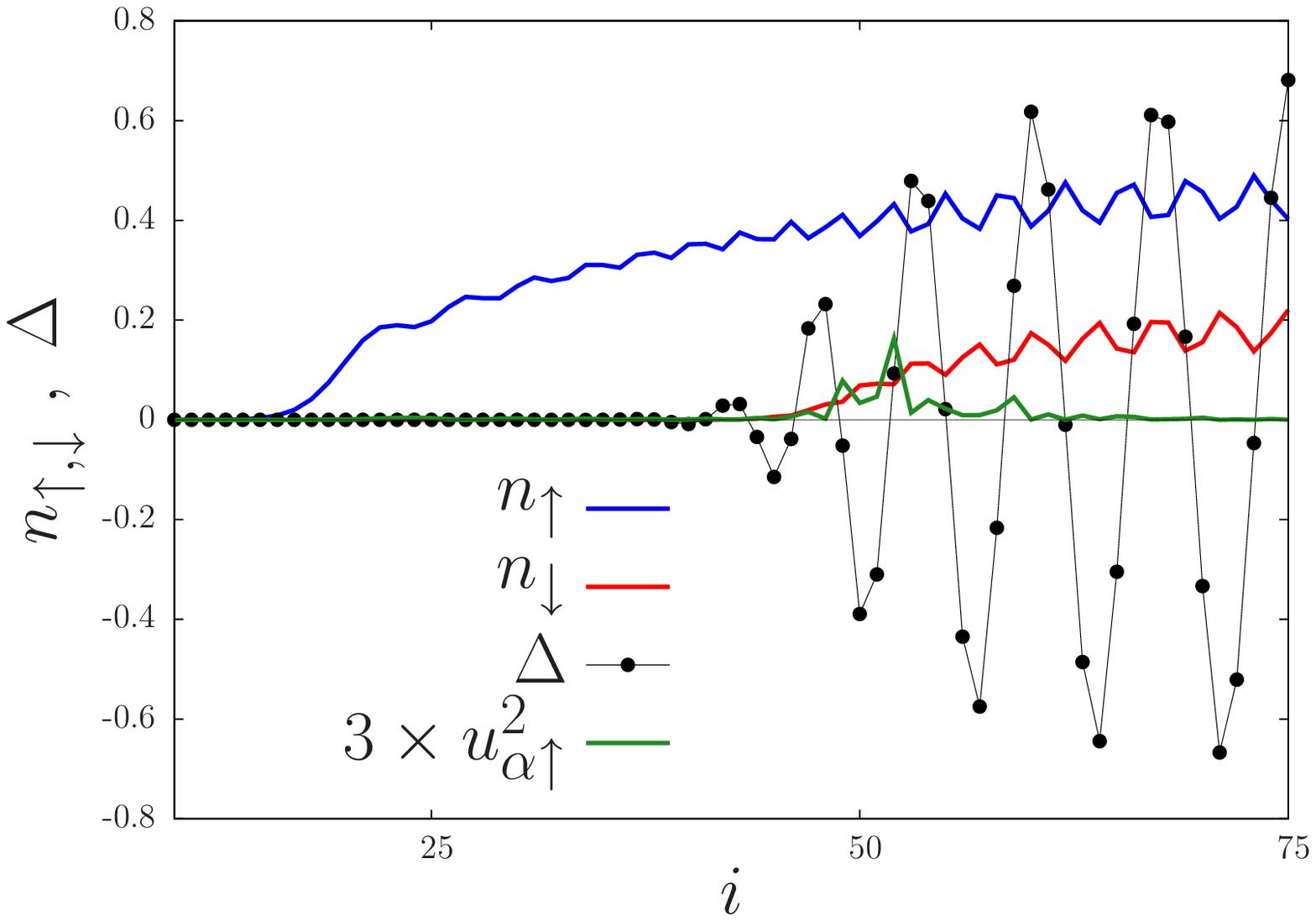}
\end{tabular}
\caption{(color online) Zero-temperature densities and order parameters $\Delta$ as a function of site position $i$ 
for polarization $P=0.23$ ($N_\ua=40,N_\da=25$) (top panel) and $P=0.7$ $(40,7)$ (bottom panel) in $L=150$ lattice sites, and in the presence of a harmonic potential.
Top panel shows $\Delta$ also at finite temperatures $T1=0.075$ and $T2=0.1$. The wavefunction $u_{123\ua}^2$ is also shown in bottom panel.} 
\label{fig:fig1}
\end{center}
\end{figure}

Typical density and order parameter profiles are shown in Fig.~\ref{fig:fig1} for the polarizations $P=0.23$ $(N_\ua=40,N_\da=25)$ and $P=0.70$ $(40,7)$ 
where the polarization is defined as $P=(N_\ua-N_\da)/(N_\ua+N_\da)$. For $P=0.23$, the order parameter is constant and the density imbalance zero in the center of trap. Surrounding that, there is an area of oscillating order parameter and finite density imbalance. For $P=0.7$, this area spans through the central part as well. At the edges, in both 
cases, the order parameter vanishes while the majority particle density is still non-zero. This kind of BCS-FFLO-Normal and FFLO-Normal shell structures are in accordance with~\cite{FFLO_papers}. For the parameters used here, the BCS part vanishes and the FFLO area reaches the trap center for polarizations $P\gtrsim 0.3$.

\begin{figure}
\begin{center}
\tabcolsep=0cm
\begin{tabular}{c}
\includegraphics[width=0.8\linewidth]{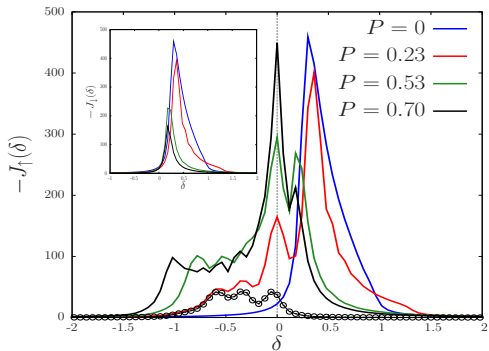}
\end{tabular}
\caption{(color online) Majority and minority (inset) RF spectra for different polarizations. The curve marked with open circles is explained in the text.} \label{fig:Fig2}
\end{center}
\end{figure}

Figure~\ref{fig:Fig2} shows the RF spectra for different polarizations. Both the minority and majority spectra have a peak located at the same, positive, detuning. This corresponds to breaking of pairs, in other words, excitations are created in the gas. The majority spectrum displays also a prominent peak at zero detuning, corresponding to unpaired majority component particles. Note that there is considerable spectral weight at negative detunings. We also plot separately the second term (proportional to $\nf(E_{\alpha\scriptscriptstyle{\ua/\da}})$) in Eq.(\ref{spectra}) for the 
polarization $P=0.23$ (open circle): this gives
almost all of the spectral signatures at negative detunings, and part of the zero detuning peak for which also the first term contributes. 
In both terms, the energy conservation relation imposes $\delta = |E_{\da\alpha}| - (\mu_\ua - \epsilon_K)$. For $\alpha = K$ this is always positive or zero, for the normal and BCS states. The BdG wavefunctions $v$ and $u$ in the interacting case are, however, not the same as in the non-interacting case and therefore the overlap integrals in Eq.(\ref{spectra}) can be non-zero also for $\alpha \neq K$. Thus, in order to obtain spectral weight at negative detunings, the interacting case wavefunctions have to be decomposed of a large set of non-interacting case wavefunctions. This may happen when the wavefunctions have to accommodate to a spatially highly non-uniform order parameter. We have analyzed the energy and eigenvalue structure of the system and found that those states that contribute to the negative detuning peaks correspond to majority particle wavefunctions that have maxima at the nodes of the order parameter. As examples, we discuss here the states $\alpha = 142$ in the  
$P=0.23$ 
case, and $\alpha=123$ for $P=0.7$. In both cases, these states correspond to negative but close to zero excitation energies, i.e. they are close to 
the Fermi surfaces. By calculating the overlap integrals, we found that the state $\alpha=142$ ($P=0.23$) spans all the way from $K=100$ to $K=140$. From the energy conservation relation one can then see that e.g.\ the $K=100$ contribution leads to spectral weight at $\delta \sim -0.6$, and the $140$ at $\delta \sim -0.3$. 
Similarly, for $P=0.7$, the state $\alpha = 123$ is decomposed of $K$ from 70 to 150, leading to contributions as far as $\delta \sim -1$. The state $\alpha=123$ is shown in Figure 1.
It resides inside the order parameter, with maxima at the nodes. Typically, states contributing to negative and zero detuning were found to be located either totally inside the order parameter, or partly outside and partly inside, and always following the node structure.
 
Some of the states discussed above are related to Andreev bound states formed at FFLO superconductor - ferromagnet interfaces~\cite{Deutscher2005a}. In semiclassical thinking, the spatially varying order parameter forms a potential affecting the unpaired majority particle excitation spectrum. In solid state systems, the Andreev bound states often lead to zero bias anomalies in tunneling experiments. RF-spectroscopy, in the quasiparticle picture as considered here, is analogous to tunneling and zero bias corresponds to zero detuning. However, part of the spectral signatures of the Andreev type states are at negative detunings due to the momentum conservation in the spectroscopy. Without momentum conservation, the overlap integrals loose their significance and transitions to all final states are allowed, then the densities of state of the final and initial state will determine the spectral weight. Since density of states typically grows for higher lying states, this 
would favour zero detuning transitions since the negative detuning transitions are to lower lying final states. The lower lying final states are often Pauli blocked in tunneling experiments.  

\begin{figure}
\begin{center}
\tabcolsep=0cm
\begin{tabular}{c}
\includegraphics[width=0.7\linewidth]{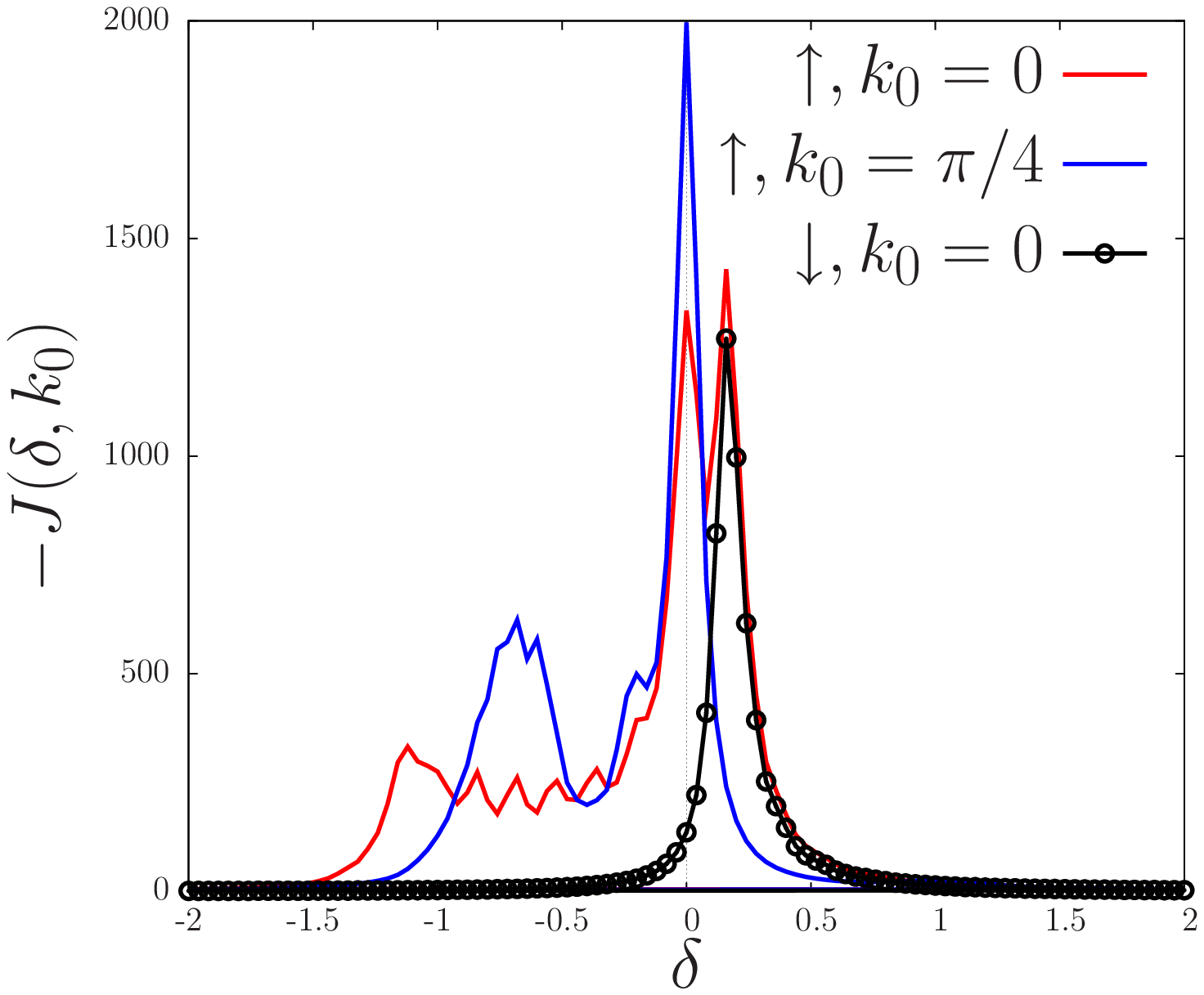}\\
\includegraphics[width=0.66\linewidth]{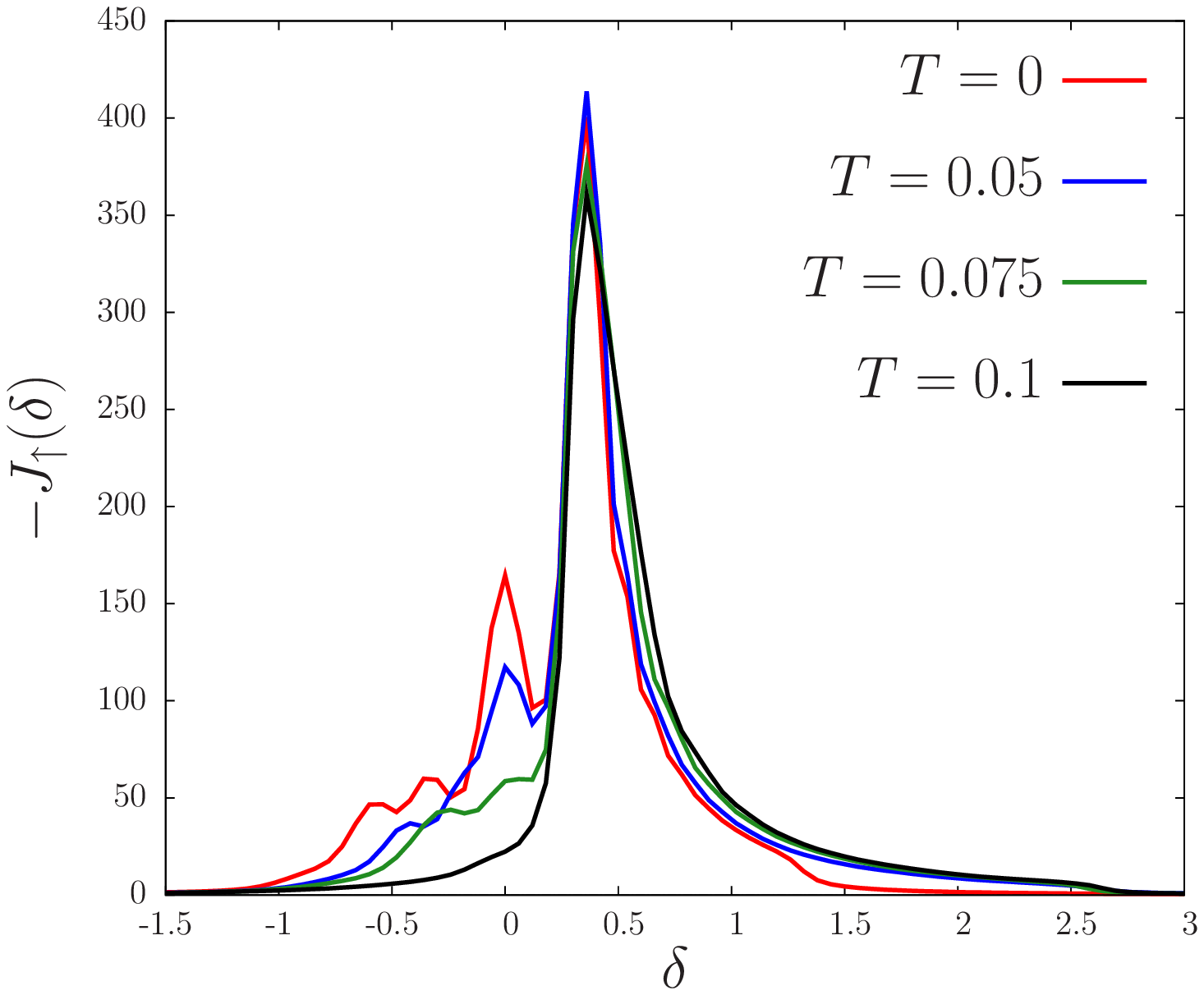}
\end{tabular}
\caption{(color online) Final state momentum resolved spectra for majority and minority components for different $k$ (top panel), for $P=0.7$. 
Majority component spectra at different temperatures for $P=0.23$ (bottom panel).} \label{fig:fig3}
\end{center}
\end{figure}

In Fig.~\ref{fig:fig3}, we show also the $k$-resolved spectra that can be observed by photoemission spectroscopy. 
The final states around the Fermi level of the majority component correspond mainly to the non-paired
particles, whereas the low momenta are involved in both pairing and non-paired excitations. 
Note that the momentum-resolved RF-spectroscopy may help to distinguish negative 
detuning contributions since peaks instead of a steady pedestal appear. 

It is now important to discuss the dependence of these negative detuning spectral features on the system size. 
We have varied the lattice size and particle number from 
$L=100 \to 150$, $N=40\to 80$ and the results are similar; from this and from the understanding of the results given above it is obvious that they are not any numerical artifact related to a finite size system. They are, however, a mesoscopic effect in the sense that the order parameter oscillations have to be in a scale that, in the semiclassical thinking, provides sufficiently tight confinement to obtain Andreev states whose spread in eigenstates of the potential is large. The spread has to be large enough to allow transitions where $\alpha$ and $K$ states are sufficiently different in energy compared to other broadenings in the experiment. Note that in~\cite{Kinnunen2006a,Mizushima2007a} 
it was shown that a peak at zero detuning in the {\it minority} spectrum is a signature of order parameter oscillations; this applies 
to large systems and for high polarizations, and is not visible in the present case. Ultracold atoms in optical lattices provide a system where 1D potentials accommodating, e.g., a few hundreds or thousands of particles can be realized. Such numbers are not essentially different from those used in our simulations, therefore we believe that spectral weight on the negative detunings should be visible in the experiment. 

Finally, although it is clear from the above discussion that the negative detuning features require the existence of a spatially non-uniform order parameter, one might still ask whether oscillations are required or is simply a smoothly varying order parameter profile sufficient? To clarify this, we calculated spectra also at finite temperatures. The disappearance of the order parameter oscillations with finite
temperature is shown in Fig.\ref{fig:fig1} and the finite $T$ spectra  in Fig.\ref{fig:fig3}, for $P=0.23$. The pedestal-type spectrum at
negative detunings disappears when the oscillations vanish. Also the zero detuning peak eventually goes down since finite temperature enables pairing even in the presence of population imbalance, i.e.\ the gas enters the finite temperature BCS phase. We can thus conclude that the prominent, pedestal-like spectrum at negative detunings, with a peak at zero, is a unique signature of an oscillating order parameter. This would be one of the most direct signatures of the FFLO state observable in any system so far.                        

In summary, we considered the density imbalanced, attractively interacting Fermi gas in a 1D optical lattice, using the mean-field BdG formalism. We propose that in such systems, RF-spectroscopy and photoemission spectroscopy can provide information not only about the pairing gap and the amount of paired/unpaired minority and majority particles, but also about the spatial structure of the order parameter. This is basically due to the momentum conservation in the process which gives an important role to wavefunction overlaps in transition probabilities and thus allows to resolve also spatial features of excitations. By analyzing spectra and eigenstates both at zero and finite temperatures, 
we have shown that significant spectral weight on negative detunings is a direct signature of the Andreev bound states or mid-gap states formed due to a spatially strongly modulated order parameter, and thereby a direct signature of the FFLO state.        
   
{\it Acknowledgments} This work was supported by the National Graduate School in Materials Physics, Academy of Finland (Project Nos.\ 213362, 217041, 217043, 210953) 
and conducted as a part of a EURYI scheme grant. See www.esf.org/euryi. M.R.B.\ acknowledges very useful discussion with M.\ Polini.
 

\end{document}